# Silicon Layer Intercalation of Centimeter-Scale, Epitaxially-Grown Monolayer Graphene on Ru(0001)

Jinhai Mao,[1] Li Huang,[1] Yi Pan,[1] Min Gao,[1] Junfeng He,[1] Haitao Zhou,[1] Haiming Guo,[1] Yuan Tian,[1] Qiang Zou,[1] Lizhi Zhang,[1] Haigang Zhang,[1] Yeliang Wang,[1] Shixuan Du,[1] Xingjiang Zhou,[1] A. H. Castro Neto,[2] Hong-Jun Gao[1,*]

[1]*Institute of Physics, Chinese Academy of Sciences, Beijing 100190, China*

[2]*Graphene Research Centre, Department of Physics, National University of Singapore, Singapore 117542*

We develop a strategy for graphene growth on Ru(0001) followed by silicon-layer intercalation that not only weakens the interaction of graphene with the metal substrate but also retains its superlative properties. This G/Si/Ru architecture, produced by silicon-layer intercalation approach (SIA), was characterized by scanning tunneling microscopy/spectroscopy and angle resolved electron photoemission spectroscopy. These experiments show high structural and electronic qualities of this new composite. The SIA allows for an atomic control of the distance between the graphene and the metal substrate that can be used as a top gate. Our results show potential for the next generation of graphene-based materials with tailored properties.

[*]Electronic mail: hjgao@iphy.ac.cn

Graphene, a two-dimensional crystal of carbon atoms arranged in a honeycomb structure, is a material with extraordinary structural and electronic properties.[1-3] Because of those properties, Graphene is being considered as a serious contender as the reference material for a post-CMOS technology.[4] The availability of high quality and large scale (wafer size) single crystalline graphene is fundamental for graphene to fulfill its promise in future electronic integrated devices.[5-7] However, nearly all grown graphene films are dominated by polycrystalline domains with randomly oriented grain boundaries.[5, 8, 9] These domains cause a dramatic deterioration of the electrical and structural properties. The elimination of those extended defects is fundamental before graphene can be used in high-end electronic applications.

In this work we propose a way that may overcome this bottleneck by using intercalation of an insulating or semiconducting layer between graphene and its metallic host. In the past, different materials have been intercalated at the interface between graphene and its substrate.[10-16] In this regard, Si plays a particularly important role, being the basis of most modern electronics technology. Although there have been suggestions in the past that graphene, due to its superlative qualities, may replace Si in the near future, it is more realistic to think that graphene can be integrated into Si technology in order to harvest the best qualities that these materials have to offer.

The Ru(0001) surface has been prepared by $Ar^+$ sputtering and annealing to the 800 $^0$C, and exposing to the oxygen at 1200 $^0$C to remove the residual carbon and then flashing to 1500 $^0$C to remove the oxide. We prepared high quality graphene by thermal decomposition of ethylene on metal substrates at high temperature. The silicon was evaporated to the graphene surface and then annealed at 500 $^0$C for 10 min. Scanning tunneling microscopy/spectroscopy (STM/STS) were performed in ultrahigh vacuum (UHV) system equipped with an Omicron UHV-LT-STM at a temperature of 5 K. The chamber was also equipped with low energy electron diffraction (LEED) and Auger electron spectroscopy (AES) to monitor the quality of the graphene. We measured the STS with lock-in technique by applying a small modulation signal, a.c. 7 mV rms at 730 Hz, to the tunneling voltage. The angle resolved electron photoemission spectroscopy (ARPES) measurements were carried out using our lab system equipped with Scienta R4000 analyzer and VUV5000UV source which gives a photon energy of Helium I at $h\upsilon = 21.218$ eV. The overall energy resolution is 10 meV and the angular resolution is 0.3 degree. The measurement was performed at 20 K in vacuum with a base pressure better than $5\times10^{-11}$ Torr.

Firstly, a highly-ordered, centimeter-scale (0.8 cm in diameter), and continuous monolayer graphene of high quality on Ru(0001)[7] was prepared for the Si layer intercalation. The procedure for silicon-layer intercalation approach (SIA) is as follows: (1) growth of monolayer graphene on Ru(0001) [Fig. 1(a)]; (2) deposition of silicon on the surface of graphene [Fig. 1(b)]; (3) annealing of the deposited Si and formation of the Si layer between the graphene and Ru [Fig. 1(c)]. We employed STM to image the surface of the samples at the three stages. For reference, the G/Ru(0001) surface structure is provided showing the moiré patterns that characterized the interaction between the Ru and graphene lattices [Fig. 1(d)]. After Si deposition on the G/Ru(0001), Si clusters can be seen on the surface of the sample [Fig. 1(e)]. After annealing at 500 $^0$C, the Si deposited clusters disappeared and the sample becomes uniform and atomically flat [Fig. 1(f)].

The structural decoupling of the graphene from the Ru surface can be seen firstly by the degree of flatness of the honeycomb lattice as compared with graphene grown directly on Ru(0001). The strong surface corrugation of the graphene on Ru(0001) disappears after the Si intercalation, as shown in Fig. 1(g). Further zooming of the STM images shows clearly the honeycomb lattice of the monolayer graphene [Fig. 1(h) and 1(i)]. We carried out DFT calculations on the interaction between Si layer and the Ru(0001) surface and have found that it is very weak with a spacing of 3 Å between graphene and the silicon layer on Ru(0001). This structural decoupling leads to the high resolution STM images on the epitaxial-grown graphene through Si intercalation.

In Fig. 2(a) ARPES measurements show the conduction ($\pi^*$) and valence ($\pi$) bands cross at the Dirac point ($E_D$) at the K point of the hexagonal Brillouin zone, with linear dispersion over a large energy range. The Dirac point is below the Fermi energy by 0.26 eV and the measured Fermi velocity (the slope of the cones in the $\Gamma$-$K$ direction) is $0.95 \times 10^6$ m/s, that is, of the same magnitude of the Fermi velocity of graphene on $SiO_2$. Notice that this result implies a charge doping of estimated $2.7 \times 10^{-3}$ electron per unit cell, i.e., $5.1 \times 10^{12}$ cm$^{-2}$. From the constant energy maps of the states at $E_F$, $E_D$ and $E_F$ -0.8 eV [Fig. 2(b) to 2(d)], we can see the hole and electron pockets, respectively. The equal energy contours in the maps which correspond to the cone structure near the $E_D$ are similar with ARPES measurements in graphene epitaxially grown in SiC.[17, 18] Another feature in the Si-intercalated graphene (SIG) is that there is no replica around the Dirac point in the constant energy mapping, which are induced by the Moiré pattern and have been observed in graphene on SiC(0001) and Ir(111). The absence of the replica around the Dirac point may indicate that the SIG is more uniform and intrinsic than graphene on SiC and Ir surfaces. Thus, graphene on the G/Si/Ru structure is decoupled from its substrate and is not subject to any significant potential that would affect the spectral properties of the Dirac quasiparticles.

Although impurities are usually considered as annoyance due to their deleterious effects on the electronic transport, they can also review important features of the electronic properties. We have selected a few samples where we can find defects (such as a vacancy) that can be used as probe of the electronic states of our samples. An STM image of the G/Si/Ru containing a local defect is shown in Fig. 3(a). It shows the superstructure around the defect, which is assigned to the inter-valley scattering of the delocalized $\pi$-electrons by the defect.[19] The fast Fourier transform (FFT) of the large area STM image with the same kinds of defect is shown in Fig. 3(b) with the (1×1) and the R3 spots, corresponding to the superstructure. The same quantum interference patterns were found in a bilayer graphene directly on Ru(0001) and graphene on SiC(0001) surface, respectively. Those were absent in monolayer graphene grown on Ru(0001) surface and in the buffer layer of SiC(0001) where the π-band is strongly perturbed.[6] Such interferences are fingerprints of π states close to the Fermi level.[20] In the upright corner of Fig. 3(b) we show one of the R3 spot in the FFT image. The fine structure in these measurements agrees well with the theoretical predictions for inter-valley scattering for chiral Dirac fermions.[21] This quantum interference at an atomic resolution demonstrates that the metallic states of the graphene in this G/Si/Ru structure are essentially decoupled from the underlying electronic states from the silicon on Ru(0001).

To investigate the electronic properties of G/Si/Ru, local tunneling conductance measurement, dI/dV, was performed over the entire area. The typical profile of LDOS for graphene is observed at low bias [Fig. 3(c)]. Similarly to graphene on $SiO_2$, the spectrum also shows a gap-like feature centered at the Fermi level and a local minimum at $V_D$= - 260 mV [Fig. 3(d)]. This gap-like feature was interpreted as a suppression of electronic tunneling to graphene states near the Fermi energy and a simultaneous enhancement of electronic tunneling at higher energies due to a phonon-mediated inelastic channel.[22] The minima at -0.26 V is associated with the Dirac point that is shifted to the electron doping side (the sample is slightly n-doped), which is in agreement with the ARPES experiments. Both of STS and the FFT-STM show the decoupling between graphene electrons and the substrate due to Si intercalation.

In order to investigate the intercalation of thicker Si layers between the graphene and the Ru(0001), we deposited more Si on top of the pre-formed monolayer-silicon intercalated graphene (G/1-Si/Ru) and then annealed the sample again. As shown in Fig. 4(a), STM images show the boundary of the bilayer Si intercalated graphene (G/2-Si/Ru) on the left and G/1-Si/Ru on the right sides of the figure. From the scale bar, the height difference is around 0.3 nm, consistent with the thickness of an added Si monolayer. Figure 4 (b) and 4(c), shows the high resolution image of graphene lattice on the G/2-Si/Ru. It can be seen that the graphene becomes uniform and much flatter than that formed on the G/1-Si/Ru, indicating even weaker interaction with the substrate. This result opens the doors for the atomic control of the distance between graphene and a metal gate (in this case Ru) by creating a G/N-Si/Ru structure (N is the number of Si layers), which would be very important for future applications.

In summary, we have demonstrated that it is possible to intercalate Si layer at the interface between epitxailly-grown, wafer-sized graphene of high quality on metal crystal surfaces (in our case Ru) and still maintain the graphene crystallinity and achieve electronic decoupling from the metal. This technique is not exclusive to Ru, but can also be used in many other metal substrates that catalyze for graphene production, such as Ni, Ir, Cu and Pt, for instance. Moreover, we have shown that it is possible to intercalate thicker layers of Si, allowing for the atomic control of the distance between graphene and the metal substrate, opening doors for controlled high doping experiments without the need of chemical doping. We would like to stress that our results indicate the possibility of incorporating graphene-based structures with Si-based materials and can be very important for future technological progress in materials science.

**Acknowledgements**

This work was supported by grants from National Science Foundation of China, National "973" project of China (No. 2009CB929103, 2011CB932700), the Chinese Academy of Sciences, and Shanghai Supercomputing Center, China. A.H.C.N. thanks DOE grant DE-FG02-08ER46512. We thank Min Ouyang for stimulating discussions and suggestions.

**Figure Captions:**

FIG. 1 (Color online). SIA on Ru(0001). (a)-(c) Schematics of the silicon intercalation process: (a) Graphene formation on Ru(0001); (b) Si deposition on Graphene/Ru(0001); (c) Annealing and Si layer intercalation. (d) STM image of graphene on Ru(0001), showing the ordered moiré pattern, 40 nm × 40 nm, 0.1 nA, -3.0 V. (e) Silicon deposition on the graphene, 20 nm × 20 nm, 0.2 nA, -2.0 V. (f) After annealing, the Si intercalation between graphene and Ru(0001), 25 nm × 25 nm, 0.1 nA, -3.0 V. (g) Zoom-in STM image of G/Si/Ru, 6 nm × 6 nm, 0.1 nA, -1.0 V. (h), (i) 3D STM images of the G/Si/Ru surface, 3 nm × 3 nm, 0.2 nA, -1.0 V.

FIG. 2 ARPES on G/Si/Ru. (a) Electronic structure of the silicon intercalated graphene along the Γ-K direction. (b)-(d) Constant energy maps at the $E_F$, $E_D$ and $E_F$ - 0.8 eV.

FIG. 3 (Color online). Silicon intercalated graphene. (a) STM image of the point defect on G/Si/Ru , 10 nm × 10 nm, 0.3 nA, -0.8 V. (b) FFT of the G/Si/Ru STM image at low voltage, 2 mV, 0.3 nA. The zoom-in FFT images show details of the inter-valley scattering spot. Each spot due to the inter-valley scattering is separated into two parts along the Γ-K direction. (c), (d) dI/dV at different bias scales. I = 0.3 nA, V = 1.5 V for (c), and I = 0.3 nA, V = 0.4 V for (d).

FIG. 4 (Color online). STM images of G/n-Si/Ru. (a) In G/1-Si/Ru region, the moiré pattern can be observed, 12 nm × 7 nm, 0.1 nA, -3.0 V. (b), (c) High resolution STM images at the G/2-Si/Ru region, showing clearly each carbon atom in the graphene lattice, 4.5 nm × 4.5 nm, 0.3 nA, -1.0 V.

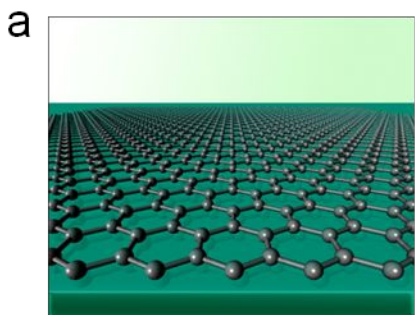 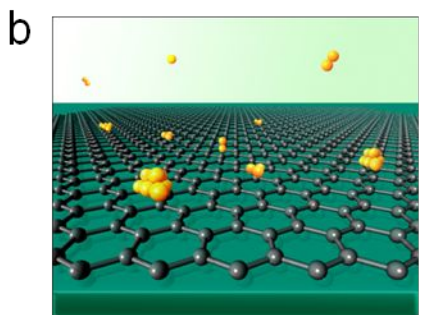 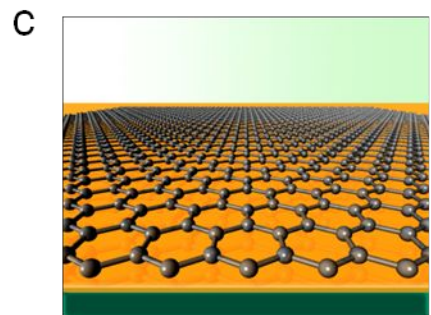

| Epitaxially-grown Graphene on Ru(0001) | Deposition of Silicon on Graphene/Ru(0001) | Silicon-Layer-Intercalation into Graphene/Ru(0001) |

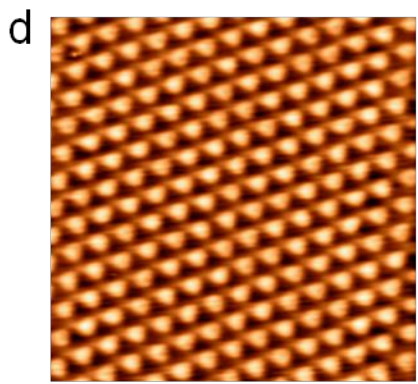 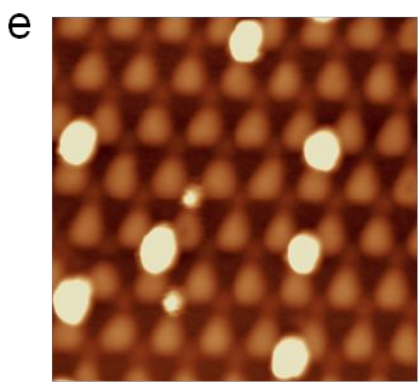 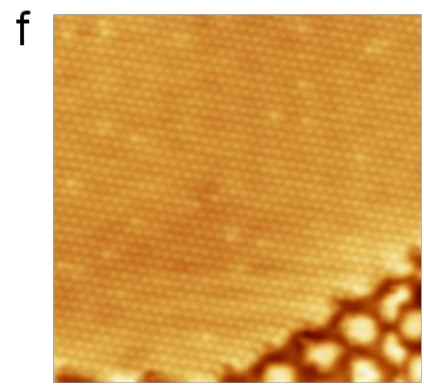

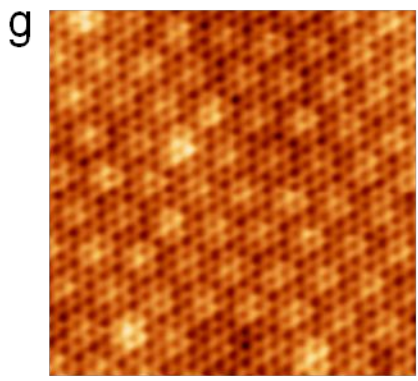 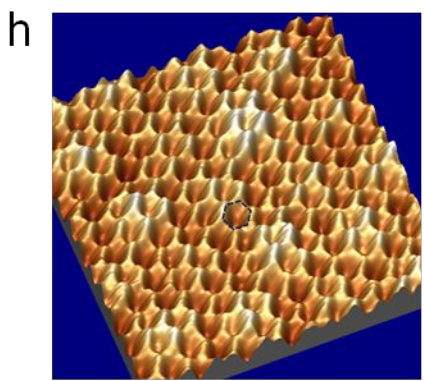 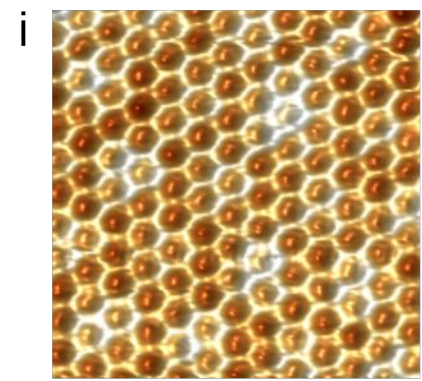

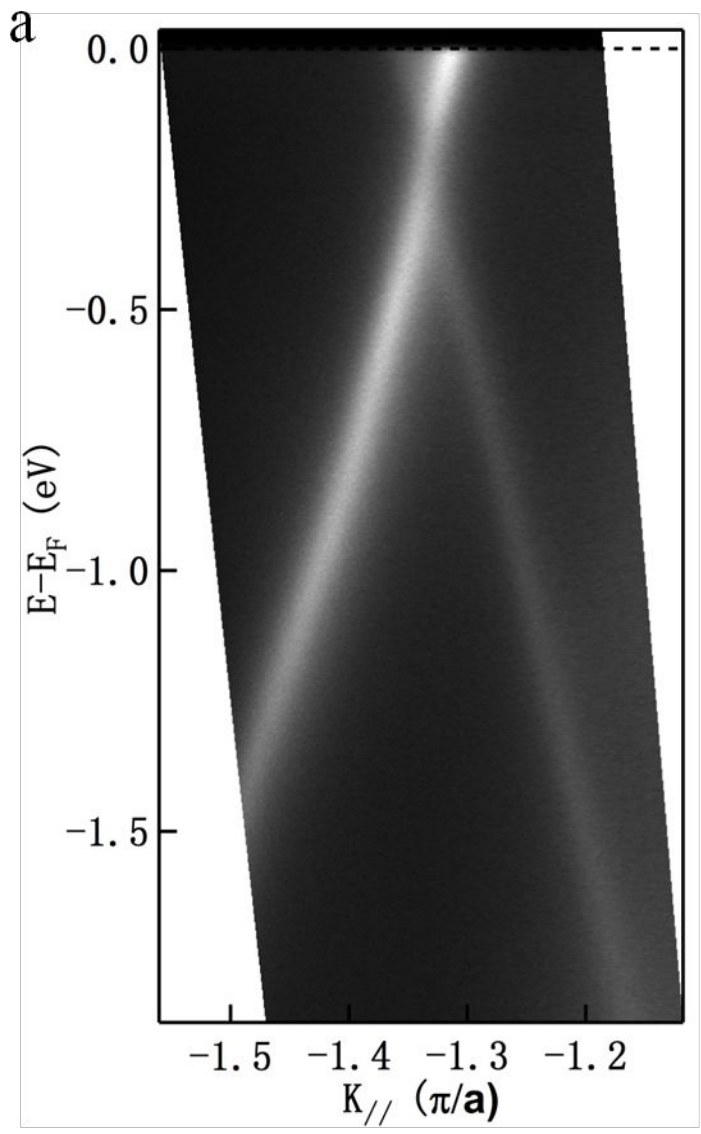
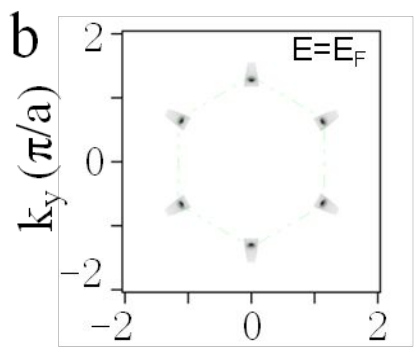
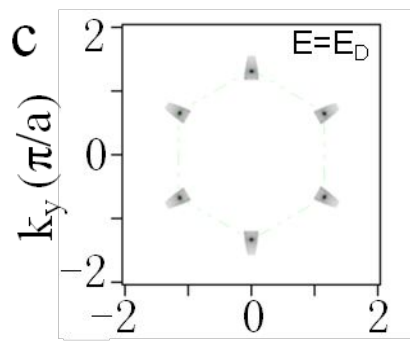
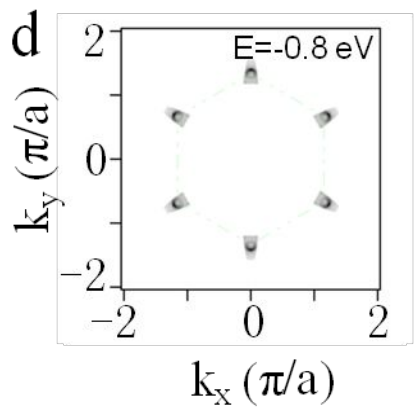

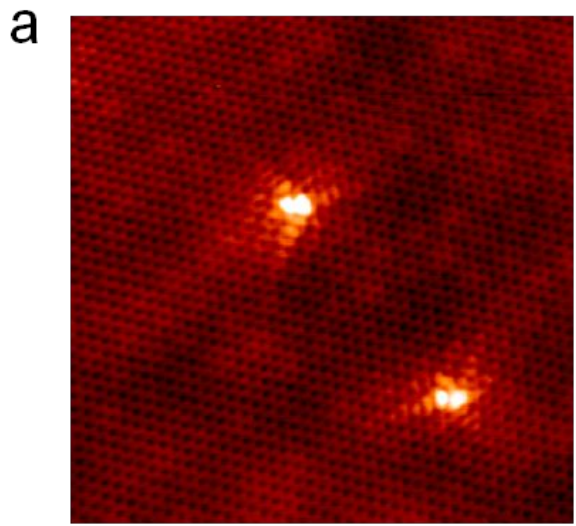
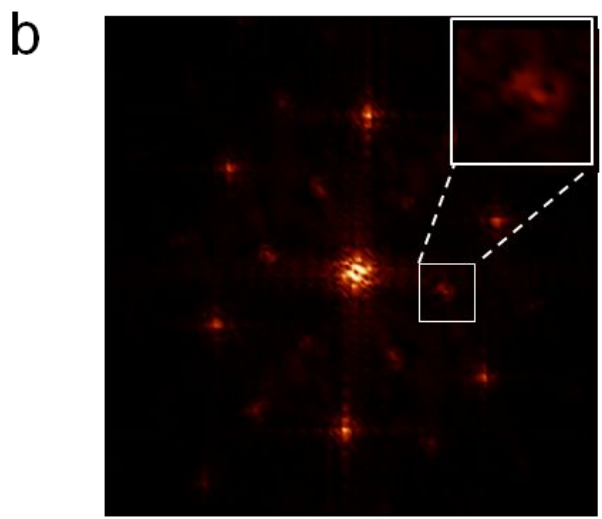
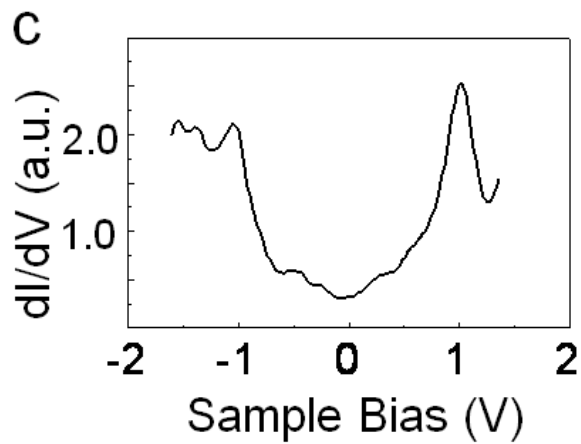
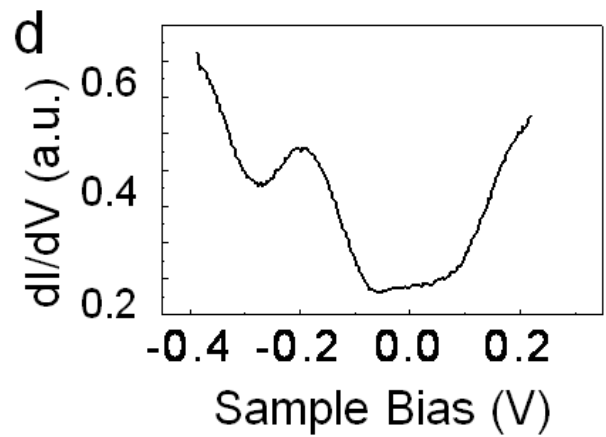

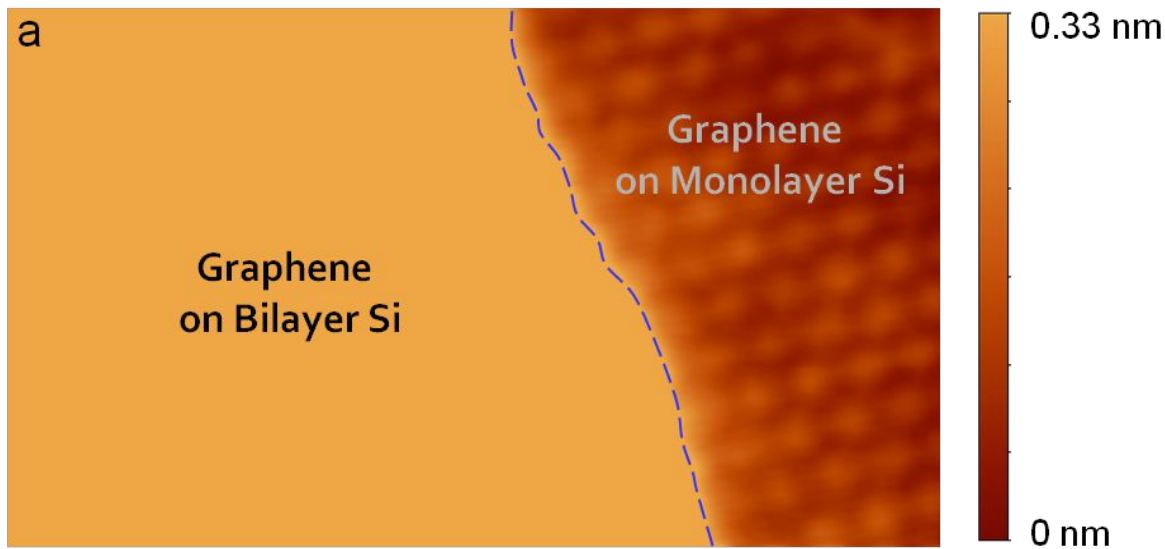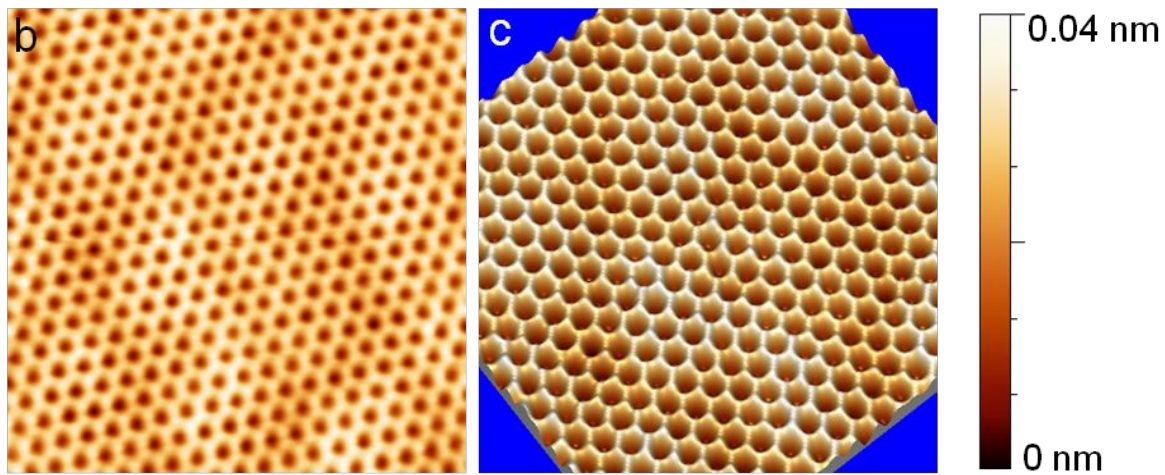